\documentclass[aps,prl,twocolumn,groupaddresses,superscriptaddress]{revtex4}%
\usepackage{graphics}
\usepackage{amsmath}
\usepackage{graphicx}
\usepackage{amsfonts}
\usepackage{amssymb}
\begin{document}
\title{Multipartite Entanglement Swapping and Mechanical Cluster States}
\author{Carlo Ottaviani}
\affiliation{Computer Science and York Centre for Quantum Technologies, University of York,
York YO10 5GH, United Kingdom}
\author{Cosmo Lupo}
\affiliation{Computer Science and York Centre for Quantum Technologies, University of York,
York YO10 5GH, United Kingdom}
\author{Alessandro Ferraro}
\affiliation{Centre for Theoretical Atomic, Molecular and Optical Physics, School of
Mathematics and Physics, Queen's University Belfast, Belfast BT7 1NN, United Kingdom}
\author{Mauro Paternostro}
\affiliation{Centre for Theoretical Atomic, Molecular and Optical Physics, School of
Mathematics and Physics, Queen's University Belfast, Belfast BT7 1NN, United Kingdom}
\author{Stefano Pirandola}
\affiliation{Computer Science and York Centre for Quantum Technologies, University of York,
York YO10 5GH, United Kingdom}

\begin{abstract}
We present a protocol for generating multipartite quantum correlations across
a quantum network with a continuous-variable architecture. An arbitrary number
of users possess two-mode entangled states, keeping one mode while sending the
other to a central relay. Here a suitable multipartite Bell detection is
performed which conditionally generates a cluster state on the retained modes.
This cluster state can be suitably manipulated by the parties and used for
tasks of quantum communication in a fully optical scenario. More
interestingly, the protocol can be used to create a purely-mechanical cluster
state starting from a supply of optomechanical systems. We show that detecting
the optical parts of optomechanical cavities may efficiently swap entanglement
into their mechanical modes, creating cluster states up to $5$ modes under
suitable cryogenic conditions.

\end{abstract}
\maketitle

\textit{Introduction.}--~~Quantum
teleportation~\cite{teleBENNETT,Braunstein98,telereview} is one of the most
important protocols in quantum information. Once two remote parties, say Alice
and Bob, have distilled maximum entanglement, they can teleport quantum
information with perfect fidelity from one location to another. In this kind
of \textquotedblleft disembodied\textquotedblright\ transport,\ the Bell
detection~\cite{Weinfurter94,Braunstein95} is one of the key operations.
Connected with quantum teleportation is the teleportation of entanglement,
also known as entanglement swapping~\cite{pirsSWAP,swap1,swap2,swap3}. Here,
Alice and Bob start with two pairs of entangled states; they then send one
part of each pair to a relay that performs Bell detection. This is a key
mechanism for quantum repeaters~\cite{Briegel,Rep2,Rep3,networkPIRS},
measurement-device independent quantum cryptography
\cite{MDI1,MDI2,CVMDIQKD,PRA-S,CVMDIQKD-reply,MDI-FS}, as well as one of tools
of a future quantum internet~\cite{Kimble,HybridINTERNET}.

In this Letter we introduce a multipartite entanglement swapping protocol for
continuous-variable (CV) systems, such as optical and/or mechanical
oscillators~\cite{BraRMP,RMP,AdessoR,sera,Paris}. We consider an arbitrary
number $N$ of users, or \textquotedblleft Bobs\textquotedblright, each having
the same identical two-mode Gaussian state $\rho_{AB}$. The $B$-modes are
kept, while the $A$-modes are sent to a central relay performing multipartite
Bell detection. The latter consists of an $N$-port interferometer, composed of
$N-1$\ cascaded beam splitters with suitable transmissivities, followed by $N$
homodyne detections. The outcomes of homodyne detection are then publicly
broadcast to all the users, which may locally apply conditional displacement operations.

The multipartite Bell detection is designed in such a way that the
output multipartite state is a symmetric Gaussian state, i.e.,
invariant under the permutation of any two Bobs. In this way, we
generate a type of Greenberger--Horne--Zeilinger (GHZ) cluster
state that the Bobs may exploit for network tasks. In the
literature, bosonic cluster states (also dubbed graph states) have
been created with different
procedures~\cite{clust1,clust2,clust3,clust4,clust5,RMP},
typically via unitary processes, e.g., by applying an
interferometer to squeezed states~\cite{vanLoock00,Yonezawa04}.
Contrary to these schemes, our strategy fully extends the approach
of Ref.~\cite{pirsSWAP}\ to a hybrid
network~\cite{hybrid1,hybrid2}, where a large supply of bipartite
states with opto-mechanical entanglement are measured in the
optical modes so that multipartite entanglement is swapped in the
mechanical modes.


Following this idea, we present an application of the proposed protocol to the
platform provided by cavity optomechanics~\cite{aspel}, which has emerged in
recent years as a promising route for the engineering of non-classical
features in mesoscopic systems. Various interesting schemes have been
suggested and, in some cases, implemented with the scope of engineering
quantum states of coupled optical and mechanical
subsystems~\cite{recenti,r2,r3,r4,r5,r6}. However, we lack a matching effort
aimed at the preparation of non-classical states of massive mechanical degrees
of freedom~\cite{houhou,vitali,v1,v2}. In this respect, the protocol put
forward here provides an interesting avenue towards the achievement of such a
tantalising goal.

\begin{figure}[ptb]
\vspace{-0.0cm}
\par
\begin{center}
\vspace{-0.7cm} \includegraphics[width=0.5\textwidth]{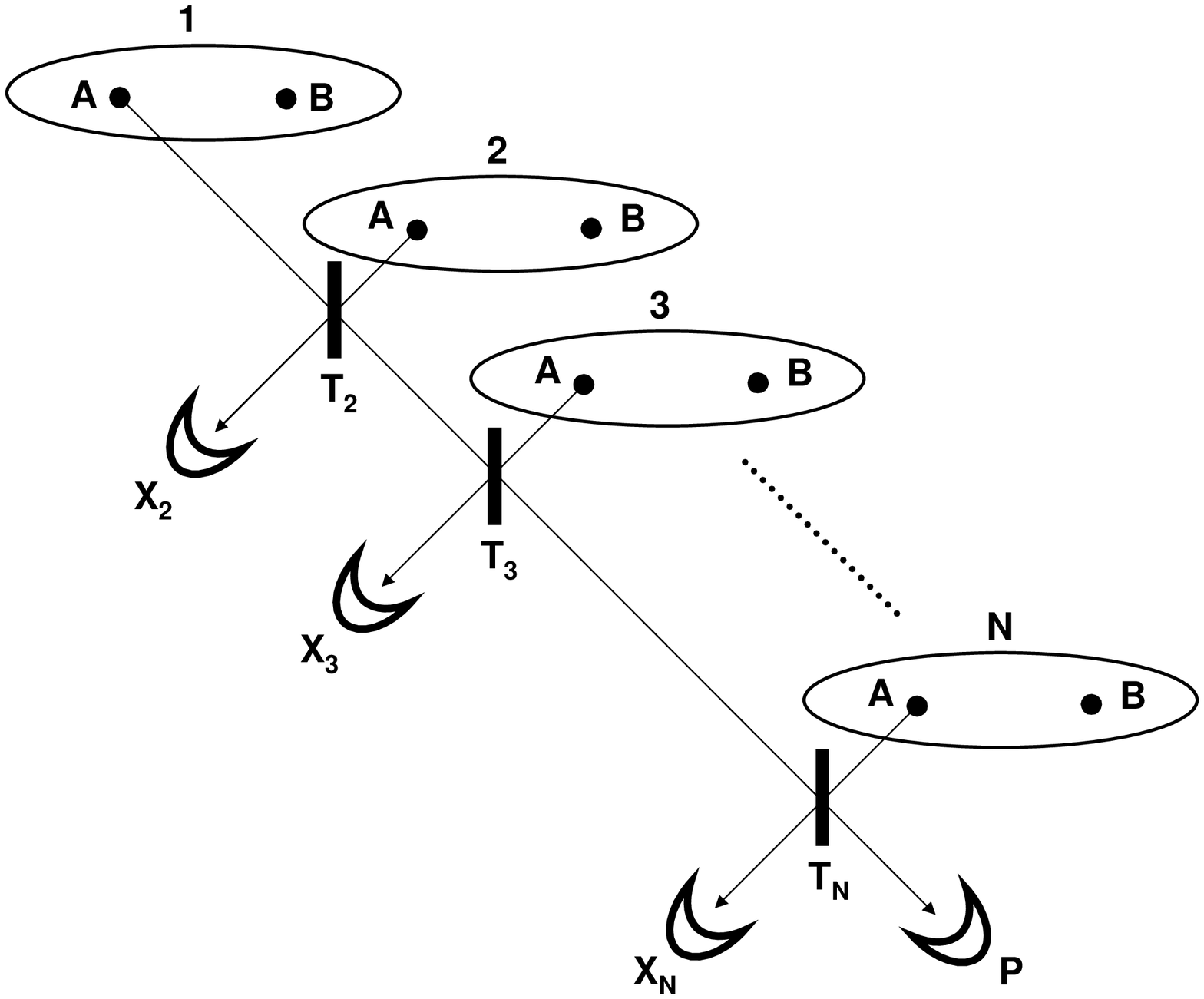}
\vspace{-0.3cm}
\end{center}
\par
\vspace{-0.5cm}\caption{Multipartite entanglement swapping. We start from $N$
independent copies of the state $\rho_{AB}$. The $A$-systems are sent to a
relay for a multipartite CV Bell detection. The latter is an interferometer
with a suitable cascade of beam splitters, followed by homodyne detections
($N-1$ in the $X$ quadratures, and a final one in the $P$ quadrature). The
outcomes $\gamma$ are broadcast to the users, so that their multipartite state
collapses into a conditional cluster state $\rho_{B_{1}\cdots B_{N}|\gamma}$.
The transmissivities $T_{k}$ of the beam splitters are chosen so that the
cluster state is invariant under permutation of the users. }%
\label{NSwap}%
\end{figure}

\textit{Multipartite entanglement swapping}.--~~Consider an ensemble of $2N$
bosonic modes which are arranged into $N$ pairs. We use the index
$k=1,\ldots,N$ for the pairs, and $A,B$ for the modes within each pair (see
Fig.~\ref{NSwap}). The whole system is described by a vector of quadratures
\begin{equation}
\vec{\xi}=(X_{1}^{A},P_{1}^{A},X_{1}^{B},P_{1}^{B},\ldots,X_{N}^{A},P_{N}%
^{A},X_{N}^{B},P_{N}^{B})^{T}~,
\end{equation}
such that $[\xi_{l},\xi_{m}]=2i\mathbf{\Omega}_{lm}$, where $l,m=1,\ldots,2N$
and $\mathbf{\Omega}$ is the symplectic form~\cite{RMP}. Within each pair $k$,
modes $A$ and $B$\ are prepared in an entangled state $\rho_{AB}$. The $A$
modes are sent to the interferometer depicted in Fig.~\ref{NSwap}, which is
defined by $N-1$ beam splitters with transmissivities $T_{k}=1-k^{-1}$ for
$k=2,\dots,N$. This interferometer transforms the input quadratures into the
output ones
\begin{align}
X_{k}  &  =\sqrt{1-k^{-1}}\left(  X_{k}^{A}-\frac{1}{k-1}\sum\limits_{i=1}%
^{k-1}X_{i}^{A}\right)  ,\label{X_k}\\
P  &  =\frac{1}{\sqrt{N}}\sum\limits_{k=1}^{N}P_{k}^{A}, \label{P_tot}%
\end{align}
which are then measured as in Fig.~\ref{NSwap}.

As a first example, consider $N$ copies of an ideal EPR state, for which we
may write \cite{BraRMP}%
\begin{equation}
P_{k}^{A}+P_{k}^{B}=0~,~X_{k}^{A}-X_{k}^{B}=0~, \label{EPRcond}%
\end{equation}
It is easy to show that the conditional state of the $B$ modes is a
multipartite CV version of the GHZ state~\cite{GHZ}, which satisfies~the
relations~\cite{BraRMP}%
\begin{align}
\sum\limits_{k=1}^{N}P_{k}^{B}  &  =0,\label{P1}\\
X_{k}^{B}-X_{k^{\prime}}^{B}  &  =0,~~\forall k,k^{\prime}=1,\ldots,N.
\label{X2}%
\end{align}
In fact, by projecting $P$ in Eq.~(\ref{P_tot}), we realize Eq.~(\ref{P1}) up
to a constant, which can be put to zero by a local displacement. In the same
way, by projecting $X_{k}$ in Eq.~(\ref{X_k}), we realize Eq.~(\ref{X2}) up to
constants~\cite{NotaREC}.

\textit{Multiswapping of Gaussian\ states}.--~~Let us compute the cluster
state generated by an input ensemble $\rho_{AB}^{\otimes N}$, where $\rho
_{AB}$ is a zero-mean Gaussian state with covariance matrix (CM) $\mathbf{V}$
in the normal form%
\begin{equation}
\mathbf{V}=\left(
\begin{array}
[c]{cc}%
x\mathbf{I} & z\mathbf{Z}\\
z\mathbf{Z} & y\mathbf{I}%
\end{array}
\right)  ,~~%
\begin{array}
[c]{l}%
\mathbf{I}=\mathrm{diag}(1,1),\\
\mathbf{Z}=\mathrm{diag}(1,-1),
\end{array}
\label{CM-GEN}%
\end{equation}
with $x$, $y$, $z$ satisfying bona-fide conditions~\cite{Bonafide}. After the
multipartite Bell detection of modes $A$ and the broadcast of the outcome
$\gamma$, the conditional cluster state $\rho_{B_{1}\cdots B_{N}|\gamma}$ of
the $B$ modes is a symmetric Gaussian state. After some algebra we compute its
CM (see also Ref.~\cite{multikey})
\begin{equation}
\mathbf{V}_{B_{1}\cdots B_{N}|\gamma}=\left(
\begin{array}
[c]{cccc}%
\mathbf{V}^{\prime} & \mathbf{C}^{\prime} & \mathbf{\cdots} & \mathbf{C}%
^{\prime}\\
\mathbf{C}^{\prime} & \mathbf{V}^{\prime} &  & \mathbf{\vdots}\\
\mathbf{\vdots} &  & \ddots & \mathbf{C}^{\prime}\\
\mathbf{C}^{\prime} & \mathbf{\cdots} & \mathbf{C}^{\prime} & \mathbf{V}%
^{\prime}%
\end{array}
\right)  ,\label{V-out}%
\end{equation}
where the blocks are given by
\begin{equation}
\mathbf{V}^{\prime}=\left(
\begin{array}
[c]{cc}%
y-\frac{N-1}{N}\frac{z^{2}}{x} & 0\\
0 & y-\frac{z^{2}}{Nx}%
\end{array}
\right)  ,~\mathbf{C}^{\prime}=\frac{z^{2}}{Nx}\mathbf{Z}.
\end{equation}

\begin{figure*}[ptbh]
\vspace{0cm} \includegraphics[width=0.9\textwidth]{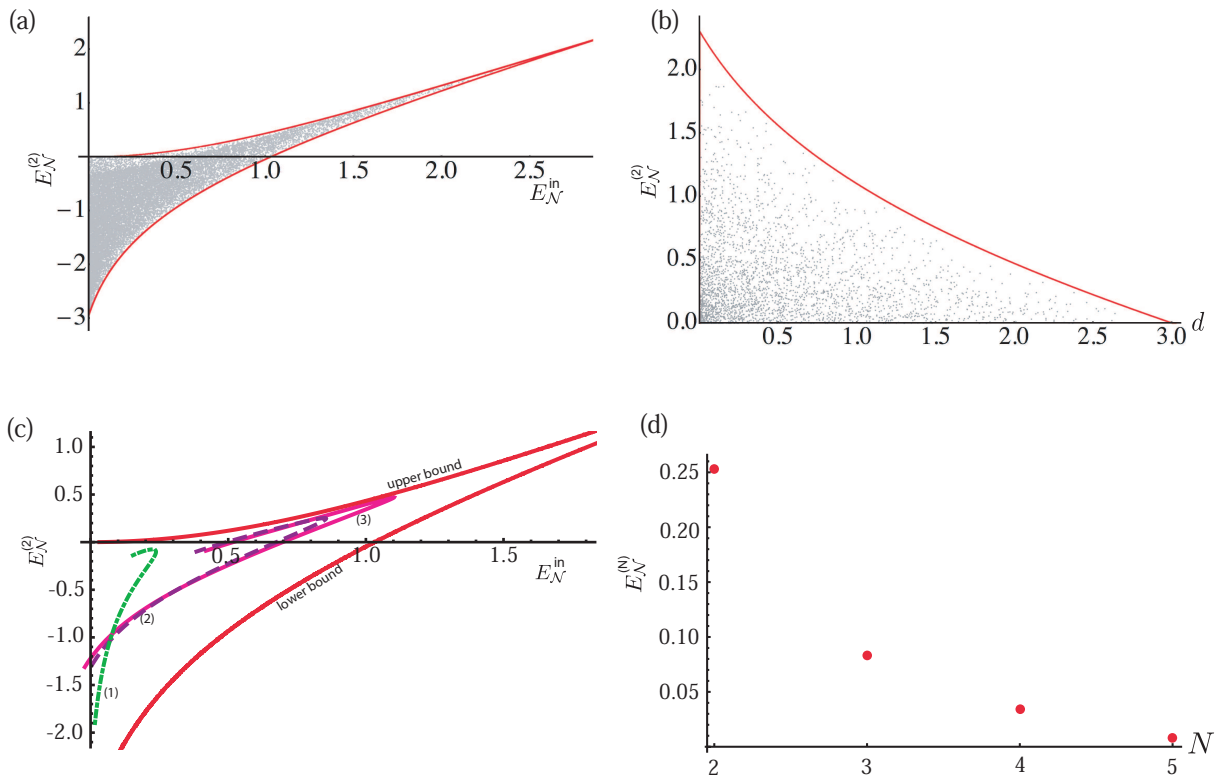}
\par
\vspace{-0.0cm}\caption{Study of the output entanglement. (a)~For $N=2$ we
plot the output log-negativity $E_{\mathcal{N}}^{(2)}$\ as a function of the
log-negativity $E_{\mathcal{N}}^{\text{in}}$ of the input Gaussian state which
is generated by random sampling. Upper and lower bounds (solid lines) are
achieved by the classes of states discussed in the main text. (b)~We show the
distribution of $E_{\mathcal{N}}^{(2)}$ as a function of the asymmetry
parameter $d$ by randomly sampling the input state. The solid line shows the
maximum achievable value. (c)~We plot the output log-negativity
$E_{\mathcal{N}}^{(2)}$ of two mechanical modes\ as a function of the input
log-negativity $E_{\mathcal{N}}^{\text{in}}$ between the optical and the
mechanical modes of two identical optomechanical systems with parameters:
$\gamma_{m}/2\pi=100$Hz, $\omega_{m}/2\pi=10$MHz, $\kappa=31.4$MHz, and
$T=0.4$mK. Each mechanical mode has mass $m=5$ng. The green dashed line $(1)$,
the dashed purple $(2)$ and solid magenta $(3)$, correspond to effective
optomechanical coupling rates of $2\pi\times4$MHz, $2\pi\times8$MHz, and
$2\pi\times8.5$MHz, respectively. The curvilinear abscissa of each line is the
detuning $\Delta\in\lbrack0,1.5\omega_{m}]$. (d)~We show the output
log-negativity $E_{\mathcal{N}}^{(N)}$ between any two modes in a cluster of
$N$ mechanical modes, for $N=2$ to $5$. Parameters as in panel (c) with an
effective optomechanical coupling strength of $2\pi\times8$MHz.}%
\label{region}%
\end{figure*}

Using Eq.~(\ref{V-out}), we may connect the log-negativity~\cite{Paris}
$E_{\mathcal{N}}^{(N)}$ between any two Bobs, $B_{i}$ and $B_{j}$, with the
log-negativity $E_{\mathcal{N}}^{\text{in}}$ of the input state $\rho_{AB}$.
For $N=2$ we may show a quasi-monotonic relation as in Fig.~\ref{region}(a),
where the gray region is generated by randomly sampling the input CM of
Eq.~(\ref{CM-GEN}) with a known parametrization~\cite{salerno}. The upper
bound is achieved by two-mode squeezed vacuum (TMSV) states, while the lower
bound corresponds to states with large asymmetry parameter $d:=(x-y)/2$. The
detrimental role of the asymmetries can also be appreciated in
Fig.~\ref{region}(b), where $E_{\mathcal{N}}^{(2)}$ is plotted versus $d$.

\textit{Cluster states in optical networks.--~~}In applications of
quantum communication, the users may be located remotely so as to
access the Bell detection via lossy optical links. Because of the
fundamental limitations affecting these links~\cite{PLOB}, the
cluster state is also degraded by loss and noise. Assume that each
Bob has a TMSV state with variance $\mu\geq 1$~\cite{RMP}. After
propagating the $A$ mode through a thermal-loss channel with
transmissivity $\eta$ and thermal noise $\omega$, the input state
$\rho_{AB}$ has CM as in Eq.~(\ref{CM-GEN}) with
$x=\eta\mu+(1-\eta)\omega$, $y=\mu$, and
$z=\sqrt{\eta}\sqrt{\mu^{2}-1}$.

From Eq.~(\ref{V-out}) we can compute the corresponding $N$-user symmetric
cluster state that is generated by the multipartite Bell detection. We find
that the log-negativity $E_{\mathcal{N}}^{(N)}$\ between any pair of Bobs
reads
\begin{equation}
E_{\mathcal{N}}^{(N)}=E_{\mathcal{N}}^{(2)}-\frac{1}{2}\ln\left(
1+\alpha\frac{N-2}{N}\right)  ,\label{formula}%
\end{equation}
where $\alpha:=\eta(\mu^{2}-1)[\eta+(1-\eta)\mu\omega]^{-1}$ and%
\begin{equation}
E_{\mathcal{N}}^{(2)}=\ln{\left[  \frac{\eta\mu+(1-\eta)\omega}{\eta
+(1-\eta)\mu\omega}\right]  }%
\end{equation}
is the log-negativity for standard swapping ($N=2$). The presence of $\alpha$
in Eq.~(\ref{formula}) shows that loss $\eta$ and noise $\omega$ destroy
entanglement more rapidly as $N$ increases~\cite{notaLOSS}.

Once the cluster state has been generated, the users may also cooperate in
such a way to concentrate the multipartite entanglement into more robust
bipartite forms. For instance, they may localize the entanglement into a pair
of users by means of quantum operations performed by all the
others~\cite{Popp}. If these operations are Gaussian, this is called Gaussian
localizable entanglement (GLE)~\cite{Fk,Fk2}. We find that the GLE
log-negativity between any pair of Bobs in the $N$-user cluster state is
\begin{equation}
E_{\mathcal{N}}^{(N,\text{GLE})}=E_{\mathcal{N}}^{(2)}-\frac{1}{2}\ln{\left(
1+\frac{N-2}{\alpha^{-1}N+2}\right)  .}%
\end{equation}

Suppose instead that the Bobs split into two groups of $N^{\prime}$ users, so
that $2N^{\prime}\leq N$. Passive unitary operations within the two groups may
map the state into a tensor product of $2N-2$ uncorrelated single-mode states
and one correlated two-mode state~\cite{Ad}. The log-negativity of the block
entanglement associated with the symmetric splitting $(N^{\prime},N^{\prime}%
)$\ of the Bobs is given by%
\begin{equation}
E_{\mathcal{N}}^{(N,N^{\prime})}=E_{\mathcal{N}}^{(2)}-\frac{1}{2}\ln{\left(
1+\alpha\frac{N-2N^{\prime}}{N}\right)  .}%
\end{equation}
Note that this is just equal to $E_{\mathcal{N}}^{(2)}$ for the
\textquotedblleft full-house\textquotedblright\ splitting $N^{\prime}=N/2$.
This is a robust concentration of entanglement because it does no longer
depend on $N$.

\textit{Generation of mechanical cluster states}.--~~We now consider the
generation of a mechanical cluster state by applying the multipartite Bell
detection to the optical parts of $N$ optomechanical systems. More precisely,
consider $N$ systems embodied by single-sided Fabry-Perot optomechanical
cavities, driven by external laser fields of suitable intensity. The
mechanical systems embody modes $B_{k}$, while the corresponding cavity fields
are the $A_{k}$'s. In a reference frame rotating at the frequency of the input
driving field, each $A_{k}-B_{k}$ interaction is modeled through the standard
radiation-pressure Hamiltonian
\begin{equation}
\hat{H}_{k}=\hbar\Delta\hat{a}_{k}^{\dag}\hat{a}_{k}+\frac{\hbar\omega_{m}}%
{2}(\hat{q}_{k}^{2}+\hat{p}_{k}^{2})-\hbar G_{0}\hat{a}_{k}^{\dag}\hat{a}%
_{k}\hat{q}+iE\hbar(\hat{a}_{k}^{\dag}-\hat{a}_{k}).
\end{equation}
Here, $\hat{q}_{k}$ and $\hat{p}_{k}$ are the dimensionless quadrature
operators of the $k^{\text{th}}$ mechanical system, $\hat{a}_{k}$ and $\hat
{a}_{k}^{\dag}$ are the ladder operators of the corresponding cavity field,
$\omega_{m}$ is the frequency of the mechanical mode (assumed to be the same
for all the mechanical systems), $G_{0}$ is the optomechanical coupling rate,
and $E$ is the amplitude of the laser drive. Finally, $\Delta$ is the laser
drive-cavity detuning.

The dynamics resulting from the Hamiltonian $\hat{H}_{k}$ is affected by the
cavity energy decay (at a rate $\kappa$) and the Brownian motion of the
mechanical oscillator (induced by the contact of each mechanical system with a
background of phonons in thermal equilibrium at temperature $T$),
characterized by the coupling strength $\gamma_{m}$. The mechanical system is
thus assumed to be prepared, prior to the optomechanical interaction, in a
thermal state at temperature $T$. The cavity is instead in a coherent state
with amplitude determined by the choice of $E$ and $\kappa$%
~\cite{optomech2,optomech}.

Under such conditions, the open dynamics at hand is well described by a set of
Langevin equations obtained considering the fluctuations around the mean
values of the operators in the problem and neglecting any non-linearity. This
is a well-established technique allowing for the gathering of information on
the quantum statistical properties of the system, as far as the fluctuations
of the operators are small compared to the mean values.
Refs.~\cite{optomech2,optomech} provide the details of the formal approach and
steps to take to derive the explicit form of the CM of the $k^{\text{th}}$
optomechanical system. From this point on, our proposed protocol for
multipartite entanglement swapping can be applied as per the previous sections.

The results are shown in Fig.~\ref{region}(c) for the case of $N=2$ and three
different choices of parameters in the optomechanical building block. The
first consideration to make is that, in line with the analysis of random
Gaussian states previously reported, the symmetry between modes $A_{k}$ and
$B_{k}$ facilitates the success of the protocol: our numerical study shows
that only for $T\ll1$, which makes the variances associated with the
fluctuation operators of the mechanical mode close to those of the cavity
field, all-mechanical entanglement might arise from the application of the
protocol. Second, such entanglement benefits of a suitably strong
optomechanical coupling rate, resulting in values that can approach the upper
boundary to the distribution in Fig.~\ref{region}(a).

Our results demonstrate the effectiveness of the proposed scheme as a method
for the achievement of all-mechanical entanglement through optical
measurements only. However, the significance of the scheme goes beyond such a
fundamental result and extends to the potential preparation of multipartite
entangled mechanical states. Indeed, we have verified that the protocol
remains successful when applied to systems of up to $N=5$ optomechanical
building blocks, as shown in Fig.~\ref{region}(d), where we report the value
of the maximum entanglement achieved as $N$ grows from 2 to 5, for the most
realistic choice of the effective optomechanical coupling strength.

\textit{Conclusions}.--~~We have introduced a protocol of multipartite
entanglement swapping for CV systems, which is based on a multipartite version
of the standard CV Bell detection. We have studied how this protocol is able
to generate an entangled cluster state in an optical lossy network, whose
entanglement can be suitably manipulated and localized by the users. Such
multipartite CV\ entangled states are useful for tasks of quantum
communication, cluster-state quantum computation~\cite{RMP}, multi-user
quantum cryptography, and distributed quantum sensing. They could also be
exploited to experimentally test gravity at the quantum level.

We have then proposed a powerful implementation of our protocol that exploits
an optomechanical interface designed to efficiently transfer entanglement onto
the mechanical modes of $N$ optomechanical cavities. Our results pave the way
towards applications for quantum technologies and networking with hybrid
architecture providing a potentially fruitful alternative to recent
experimental demonstration of all-mechanical
entanglement~\cite{Groeblacher,new}.

\textit{Acknowledgements}.-- SP, CO and CL acknowledge support
from the EPSRC via the `UK Quantum Communications HUB' (Grant no.
EP/M013472/1) and the Innovation Fund Denmark (Qubiz project). AF
is supported by the UK EPSRC (grant EP/N508664/1). MP acknowledge
the DfE-SFI Investigator Programme (grant 15/IA/2864), the Royal
Society and the COST Action CA15220 "Quantum Technologies in
Space".


\begin{thebibliography}{99}
\bibitem {NotaREC}In fact, according to Eqs.~(\ref{X_k})-(\ref{P_tot}), the
first projection onto $X_{2}$ gives (up to a constant) $X_{1}^{A}-X_{2}^{A}%
=0$. By using the latter into the expression of $X_{3}$, we derive $\sqrt
{6}X_{3}=2X_{1}^{A}-2X_{3}^{A}$. The projection onto $X_{3}$ gives therefore
the second condition $X_{1}^{A}-X_{3}^{A}=0$, and so on.
\bibitem {notaLOSS}In lossless and noiseless conditions ($\eta=\omega=1$), for
large $\mu$ we may write $E_{\mathcal{N}}^{(N)}=-\frac{1}{2}\ln\left(
1-2/N\right)  $, which is always positive but scales as $O(1/N)$.
\end{thebibliography}
\end{document}